\documentclass[aps,prl,reprint,groupedaddress,twocolumn,superscriptaddress]{revtex4-1}
\usepackage{graphicx}     
\usepackage{subfigure}
\usepackage{dcolumn}     
\usepackage{bm}              
\usepackage{hyperref}     
\usepackage{amsmath}   
\usepackage{amssymb}
\usepackage{tabularx}
\usepackage{wasysym}
\usepackage{float}
\usepackage{color}
\usepackage[table]{xcolor}
\usepackage{array}
\usepackage{multirow}
\usepackage{makecell}
\usepackage[version=3]{mhchem} 
\usepackage{float}
\usepackage{lipsum}
\usepackage{ulem}

\usepackage{verbatim}    
 
\makeatletter
\newcommand{\colorcaption}[2][]{%
  \begingroup%
  \renewcommand{\@caption@fignum@sep}{ (color online). }%
  \caption[#1]{#2}%
  \endgroup%
  }
\makeatother

\makeatletter
\newcommand*{\rom}[1]{\expandafter\@slowromancap\romannumeral #1@}
\makeatother

\begin{document}
\title{In-plane critical magnetic fields in magic-angle twisted trilayer graphene}
\author{Wei Qin}
\author{Allan H. MacDonald}
\affiliation{Department of Physics, The University of Texas at Austin, Austin, Texas 78712,USA}
\email{macd@physics.utexas.edu}
\date{\today}

\begin{abstract}
It has recently been shown \cite{Cao:2021aa} that superconductivity in magic-angle twisted trilayer graphene
survives to in-plane magnetic fields that are well in excess of the Pauli limit, and much stronger than the in-plane critical magnetic fields of magic-angle twisted bilayer graphene. The difference is surprising because 
twisted bilayers and trilayers both support the magic-angle flat bands thought to be the fountainhead of twisted 
graphene superconductivity.  We show here that the difference in critical magnetic fields
can be traced to a $\mathcal{C}_2 \mathcal{M}_{h}$ symmetry in trilayers that survives
in-plane magnetic fields, and also relative displacements between top and bottom layers
that are not under experimental control at present.
An gate electric field breaks the $\mathcal{C}_2 \mathcal{M}_{h}$ symmetry and therefore 
limits the in-plane critical magnetic field. 
\end{abstract}

\maketitle

\textit{Introduction--} 
Superconductivity has been observed in magic-angle twisted bilayer graphene (MATBG) over a broad range of flat-band fillings and electrical screening environments \cite{Cao:2018aa,Yankowitz:2019aa,Lu:2019aa,Balents:2020aa,Stepanov:2020aa,Saito:2020aa,Stepanov:2020ab,Liu:2021aa}.
The underlying mechanism responsible for superconductivity in MATBG remains under active debate;
both strong electron-electron interaction driven unconventional superconductivity \cite{Guo:2018aa,Dodaro:2018aa,Liu:2018aa,Isobe:2018aa,Fidrysiak:2018aa,Kennes:2018aa,Sherkunov:2018aa,Xie:2019aa,Kerelsky:2019aa,Choi:2019aa,Jiang:2019aa,Gonzalez:2019aa,Roy:2019aa,Huang:2019aa,Ray:2019aa,You:2019aa,Khalaf:2020aa,Qin:2021aa,Cea:2021aa}, and electron-phonon interaction mediated conventional superconductivity \cite{Wu:2018aa,Choi:2018aa,Lian:2019aa,Angeli:2019aa,Lewandowski:2021aa} have been considered theoretically.
The recent confirmation of superconductivity in magic-angle twisted trilayer graphene (MATTG) \cite{Park:2021aa,Hao:2021aa,Tsai:2019aa}
represents an important advance because MATTG and MATBG share nearly identical flat bands at twist
angles that differ by a factor of $\sqrt{2}$ \cite{Khalaf:2019aa,Mora:2019aa,Carr:2020aa,Zhu:2020aa,Lei:2020aa}, but also have important differences.  In particular, the trilayer 
hosts both even parity flat bands and odd parity dispersive bands (absent in the bilayer case)
that can be mixed by a mirror-symmetry-breaking electric displacement fields \cite{Park:2021aa,Hao:2021aa}.

Since the densities of states of MATBG and MATTG are both dominated near neutrality by magic-angle flat bands,
it is not surprising that the two systems share many properties, including similar patterns of broken flavor symmetries,
and a strong superconducting dome between moir\'e filling factors $\nu =-2$ and $-3$ \cite{Park:2021aa,Hao:2021aa}. 
It is therefore remarkable that MATBG and MATTG superconductors differ qualitatively in their response to in-plane magnetic fields.  Whereas the in-plane critical magnetic field is 
comparable with the Pauli limit in MATBG \cite{Cao:2021ab},  this limit is exceeded by nearly a factor of three 
in MATTG \cite{Cao:2021aa}.  We show below that this surprising observation can be explained if 
we assume that both systems have valley-singlet spin-triplet Cooper pairs.

Generally speaking, superconductivity is suppressed by magnetic fields because they break
the time-reversal symmetry that guarantees degeneracy of the electron pairs that combine to 
form Cooper-pair bound states.  For example, if the Cooper pairs in MATBG and MATTG were spin-singlets, 
Zeeman splitting $\Delta_z$ of opposite spins would suppress superconductivity when the Pauli limit $\Delta_z \approx 1.75 k_B T_c $
is exceeded \cite{Chandrasekhar:1962aa,Clogston:1962aa}.  In this work we take the view that superconductivity must be nearly identical in MATBG and MATTG.
Since the Pauli limit is exceeded in MATTG \cite{Cao:2021aa}, we conclude that the Cooper pairs must be spin triplets not only in 
trilayers but also in bilayers.  We will see that this view nevertheless provides a natural explanation 
for the difference in-plane critical fields.
Indeed there is some evidence \cite{Xie:2020ab,Potasz:2021aa,Fischer:2021aa} 
that in both systems the state from which superconductivity emerges is a spin 
polarized ferromagnet, leaving spin-triplets as the only pairing possibility. 

\begin{figure}
 \centering
\includegraphics[width=0.9\columnwidth]{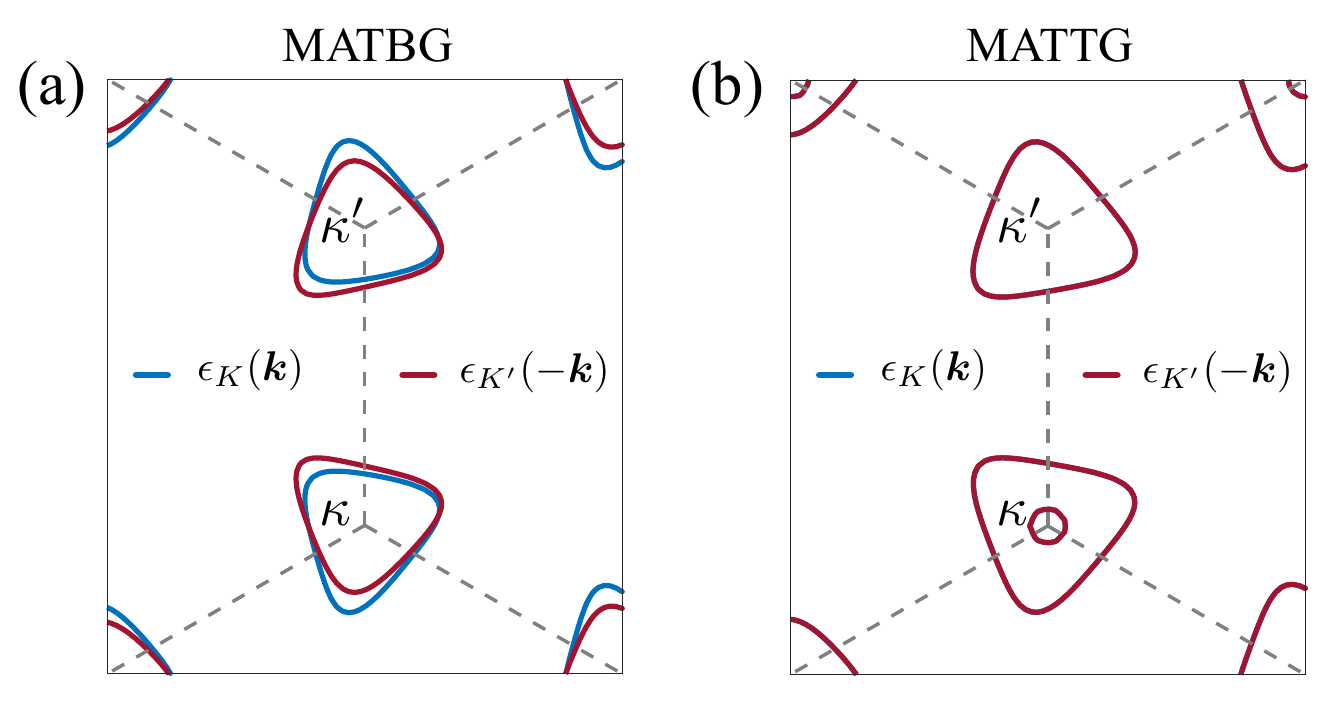}
 \caption{Nesting between $\epsilon_{K}(\bm{k})$ and 
 $\epsilon_{K'}(-\bm{k})$ Fermi surfaces for in-plane magnetic field $B_x=2$ T and moir\'e filling factor 
 $\nu=-2.4$  (See main text for model details) in   
 (a) MATBG and (b) MATTG. In bilayers (a), $\epsilon_{K}(\bm{k}) \neq \epsilon_{K'}(-\bm{k})$ due to broken 
 $\mathcal{T}$ symmetry. In trilayers (b), $\epsilon_{K}(\bm{k}) = \epsilon_{K'}(-\bm{k})$ survives
 (and one Fermi surface is hidden) because this degeneracy is also guaranteed by $\mathcal{C}_2\mathcal{M}_h$ symmetry, which survives as summarized  in Table.~\ref{tab:table1}.}
 \label{fig:figure1}
\end{figure}

In the absence of a magnetic field, time reversal ($\mathcal{T}$) symmetry guarantees that band states with 
opposite momentum in opposite valleys are degenerate.  These are the states that pair in valley-singlet superconductors.
In-plane magnetic fields break $\mathcal{T}$ symmetry in both bilayers and trilayer.  
In bilayers this produces 
an energy splitting that suppresses superconductivity \cite{Cao:2021ab,Wu:2019aa}. 
In trilayers, however, both time-reversal and mirror ($\mathcal{T} \mathcal{M}_h$) and twofold rotation and mirror ($\mathcal{C}_2 \mathcal{M}_h$) symmetries survive (see Table ~\ref{tab:table1}) and, as we explain below,
independently guarantee the degeneracy that supports valley-singlet superconductivity.
In Fig.~\ref{fig:figure1}, we illustrate this qualitative difference by 
comparing typical Fermi surfaces of MATBG and MATTG at finite in-plane magnetic field.

\begin{table}[h]
\caption{Symmetries of MATTG models with different attributes (see main text). 
We distinguish symmetry operations that map electronic
states between valleys (intervalley) from those that preserve valley (intravalley).  
$\mathcal{T} \mathcal{M}_h$ and $\mathcal{C}_2 \mathcal{M}_h$ symmetries are equivalent 
when intravalley $\mathcal{C}_2\mathcal{T}$ symmetry is present.  Layer energy refers to the difference between $\pi$-orbital energies on interior and exterior layers. Sublattice refers to sublattice polarization within layers.}
\label{tab:table1} 
\begin{ruledtabular}
\begin{tabular}{ l | c c c c | c c c}
 & \multicolumn{4}{c|}{Intravalley} & \multicolumn{3}{c}{Intervalley} \\
 & $\mathcal{C}_{3}$ & $\mathcal{M}_h$ & $\mathcal{C}_{2}\mathcal{T}$ & $\mathcal{C}_2 \mathcal{T} \mathcal{M}_h$  & $\mathcal{T}$ &$\mathcal{T} \mathcal{M}_h$  & $\mathcal{C}_2 \mathcal{M}_h$ \\
     \hline
Layer energy& \checkmark & \checkmark  & \checkmark &  \checkmark & \checkmark & \checkmark  & \checkmark \\
Gate field  & \checkmark & $\times$  & \checkmark &  $\times$  & \checkmark & $\times$ & $\times$ \\
Lateral shift  & $\times$ & $\times$ & $\times$  &  \checkmark &\checkmark & $\times$ & \checkmark \\
In-plane $\bm{B}_{||}$ &  $\times$  &$\times$  &  \checkmark  & $\times$ & $\times$ & \checkmark &\checkmark \\
Sublattice  & \checkmark & \checkmark & $\times$  & $\times$  &\checkmark & \checkmark  & $\times$ \\ 
\end{tabular}
\end{ruledtabular}
\end{table}

Below we first confirm this symmetry argument by 
performing mean-field calculations of in-plane magnetic field dependent
superconducting critical temperatures, using continuum model band structures
and a phenomenological attractive interaction.  
We then study the influence on the in-plane critical magnetic field
of gate electric fields,
which can easily be tuned experimentally, 
and lateral shifts of the top or bottom graphene layer, which may occur accidentally.
We find that the former breaks both $\mathcal{T} \mathcal{M}_h$ and $\mathcal{C}_2\mathcal{M}_h$ symmetries, 
leading to a reduced in-plane critical magnetic field, while the later preserves $\mathcal{C}_2\mathcal{M}_h$ 
symmetry thereby retaining state degeneracy and robust superconductivity. 

\textit{Band structure and symmetries ---}
An approximate single-particle band structure model of MATTG can be constructed 
by generalizing the Bistritzer-MacDonald MATBG model \cite{Bistritzer:2011aa}. 
In the absence of gate electric fields and lateral shifts, the $K$-valley projected Hamiltonian  
\begin{equation}
\mathcal{H}_{K}=\begin{bmatrix}
h_{\theta/2}(\bm{k}) & T(\bm{r}) &0\\
T^{\dagger}(\bm{r}) & h_{-\theta/2}(\bm{k}) &T^{\dagger}(\bm{r})  \\
0 & T(\bm{r})  & h_{\theta/2}(\bm{k})
\end{bmatrix},
\label{eq:TTG_K}
\end{equation}
where $h_{\theta}(\bm{k}) =  e^{i\frac{i\theta}{2} \sigma_z} (v_D\bm{k} \cdot \bm{\sigma}) e^{-i\frac{\theta}{2} \sigma_z}$, $v_D$ is the Dirac velocity of isolated monolayer graphene, and $\bm{\sigma} = (\sigma_x,\sigma_y)$ are Pauli matrices acting on sublattice. The interlayer tunneling matrix $T(\bm{r}) = \sum_{n=1}^3 T_n e^{i\bm{q}_n \cdot \bm{r}}$, where $T_{n+1} = t_{\text{AA}}\sigma_0 + t_{\text{AB}} [\cos(n \phi)\sigma_x+\sin(n\phi)\sigma_y] $ and $\phi = 2\pi/3$. It follows from mirror symmetry that $\mathcal{H}_{K}$
can be written in a representation of decoupled even-parity and odd-parity states \cite{Zhu:2020aa,Carr:2020aa}.
From Eq.~(\ref{eq:TTG_K}), we see that the trilayer's even parity subspace Hamiltonian maps to that of a bilayer, with the even parity
combination of the trilayer's outer layers playing the role of a single layer.  Because two different tunneling terms couple the inside and outside layers, the effective bilayer tunneling amplitude, 
(and therefore the magic twist angle, the flat band width, and the density of states) 
increases by a factor of $\sqrt{2}$.   The odd parity band, also formed from the outside layers,  
is identical to that of an isolated graphene layer and strongly dispersive.

Although Eq.~(\ref{eq:TTG_K}) captures the essential properties of MATTG, the modifications that we classify
in Table ~\ref{tab:table1} can be important.
First, since the chemical environments of the middle and outer graphene layers in MATTG are different, there is a layer energy difference 
between them \cite{Lei:2020aa}, which does not break any symmetry.  Other modifications break one or more of the  
$\mathcal{T}$, $\mathcal{C}_3$, $\mathcal{M}_h$, $\mathcal{C}_2\mathcal{T}$, and $\mathcal{C}_2 \mathcal{M}_h \mathcal{T}$ symmetries of 
undisturbed trilayers. For example, the electric fields routinely applied 
using gates break $\mathcal{M}_h$ symmetry and hybridize the dispersive Dirac bands and flat bands. 
Because $\mathcal{C}_3$ and $\mathcal{C}_2 \mathcal{T}$ symmetries survive this perturbation, 
gapless Dirac cones remain at the $\kappa$ and $\kappa'$ points of the moir\'e Brillouin zone (MBZ) \cite{Carr:2020aa,Zhu:2020aa,Lei:2020aa}. 
Additionally, relative translational shifts between the top and bottom graphene layers, which are not under experimental control at present,
can occur even though first-principles calculations show that
the mirror symmetric configuration is energetically most stable \cite{Carr:2020aa}. Lateral shifts
can be captured by adding phase factors to $T(\bm{r})$ \cite{Bistritzer:2011aa} and break $\mathcal{C}_3$, 
$\mathcal{M}_h$ and $\mathcal{C}_2 \mathcal{T}$ symmetries. 
The gapless Dirac cones nevertheless remain because they are protected by 
$ \mathcal{C}_2 \mathcal{T} \mathcal{M}_h$ symmetry, and simply 
move away from $\kappa$ and $\kappa'$. 
Table~\ref{tab:table1} summarizes these symmetry considerations.  None of the 
model attributes discussed above generate a band gap at any energy.

\textit{In-plane magnetic fields ---} 
The application of in-plane magnetic field $\bm{B}_{||}$ to MATBG or MATTG induces a layer-dependent gauge field $\bm{A}_{l} = \bm{B}_{||} \times \bm{z}_l$ \cite{Wu:2019aa,Kwan:2020aa}. For MATBG, the gauge field shifts the momenta of electrons in the the top and bottom graphene layers along opposite directions so that 
\begin{equation}
\mathcal{H}_{K}(\bm{k})=\begin{bmatrix}
h_{\theta/2}(\bm{k}+\bm{p}/2) & T(\bm{r}) \\
T^{\dagger}(\bm{r}) &h_{-\theta/2}(\bm{k}-\bm{p}/2)
\end{bmatrix},
\label{eq:TBG_K}
\end{equation}
where $\bm{p} = (\pi d/\Phi_0)\, (B_y,-B_x)$, $d$ is the interlayer distance, $\Phi_0$ denotes the magnetic flux quantum, and we have placed $z = 0$ at the center of the two layers. 
Since the momenta of the electrons from the same layer but in opposite valleys are shifted along the same direction, 
$\mathcal{T}$ symmetry is broken, and $\delta E \equiv \epsilon_{K'}(\bm{-k}) - \epsilon_{K}(\bm{k}) \neq 0$ as illustrated in Fig.~\ref{fig:figure1}(a),
suppressing contributions to valley singlet ladder sums.  
Therefore, the application of $\bm{B}_{||}$ leads to a reduction on the superconducting critical temperature $T_c$.
The critical magnetic field $B_c$ at which $T_c$ is driven to zero is reached 
when $ \delta E  \sim \Delta $, where $\Delta$ is the gap in the 
absence of the magnetic field.

In the MATTG case, choosing $z=0$ in the middle layer,
the gauge field shifts momenta of top and bottom layer electrons  
along opposite directions, while leaving the middle graphene layer unaffected. 
The $K$-valley Hamiltonian
\begin{equation}
\mathcal{H}_{K}(\bm{k})=\begin{bmatrix}
h_{\theta/2}(\bm{k}+\bm{p}) & T(\bm{r}) &0\\
T^{\dagger}(\bm{r}) & h_{-\theta/2}(\bm{k}) &T^{\dagger}(\bm{r})  \\
0 & T(\bm{r})  & h_{\theta/2}(\bm{k}-\bm{p})
\end{bmatrix},
\label{eq:TTG_K_B}
\end{equation}
has broken $\mathcal{C}_3$ and $\mathcal{M}_{h}$ symmetries.
Note that the sign of $\bm{p}$ is changed by a miror operation.
Although $\mathcal{T}$ symmetry is broken, as summarized in Table~\ref{tab:table1}, 
the two valleys can still be mapped to each other by either the combined $\mathcal{C}_2 \mathcal{M}_{h}$ or $\mathcal{T} \mathcal{M}_{h}$ symmetry. In MATTG,  
$\epsilon_{K'}(\bm{-k}) = \epsilon_{K}(\bm{k})$ even in the presence of $\bm{B}_{||}$. 
This property explains the perfect intervalley Fermi surface nesting as shown in Fig.~\ref{fig:figure1}(b). 
Because the quasiparticle pairs from which the Cooper pairs are formed retain degeneracy there is no obvious mechanism to 
suppress superconductivity.  Indeed, numerical model calculations summarized below suggest 
that superconductivity in MATTG can survive at extremely large values of $\bm{B}_{||}$. 
On the other hand simultaneous breaking of $\mathcal{C}_2 \mathcal{M}_{h}$ and  $\mathcal{T} \mathcal{M}_{h}$ symmetries by 
an gate electric field, lifts the pairing degeneracy and leads to a reduced in-plane critical magnetic field.

\textit{Numerical model calculations ---} Superconductivity occurs in MATBG and MATTG when each is close to its 
magic rotation angle.  We therefore compare the two systems with $\theta_{\text{TBG}} = 1.1^{\circ}$ 
and $\theta_{\text{TTG}} = \sqrt{2} \times 1.1^{\circ}$.
First principles calculations show that the moir\'e patterns of both systems distort to 
avoid high energy local AA stacking, leading to $t_{\text{AA}}/t_{\text{AB}}<1$ \cite{Carr:2019aa,Carr:2020aa}. 
Here we take $t_{\text{AA}}/t_{\text{AB}} = 0.7$ for both systems. 
Similarly, particle-hole asymmetric behavior has also been observed in the transport measurements of MATTG \cite{Park:2021aa,Hao:2021aa}, and becomes dramatic near the magic angle. This property can be modeled by including a nonlocal momentum-dependent correction to
interlayer tunneling \cite{Xie:2020ab}.  We choose $d t_{\text{AA}}/dk = d t_{\text{AB}}/dk = -0.1$ in this study. 
The nonlocal interlayer tunneling also increases the energy difference between the dispersive Dirac bands and the flat bands of MATTG. 
Below, the moir\'e filling factor is fixed at $\nu = -2.4$, where the critical temperature $T_c$ is commonly maximized
in both MATBG and MATTG superconductors \cite{Cao:2018aa,Yankowitz:2019aa,Lu:2019aa,Park:2021aa,Hao:2021aa}. As discussed in the introduction, we take the view that the normal 
state at $\nu = -2.4$ is spin polarized due to flavor symmetry breaking \cite{Potasz:2021aa}.
In the model we study,
the Fermi surfaces are located around the $\kappa$ and $\kappa'$ points of MBZ as shown in Fig.~\ref{fig:figure1}.
This detail is not yet established experimentally. 

We perform mean-field Bogoliubov-de Gennes calculations that account for in-plane magnetic fields to determine critical temperatures and 
fields.  Our calculations employ a model interaction Hamiltonian
\begin{equation}
\begin{aligned}
H_{int} &= U \sum_{l \sigma} \int d\bm{r} \psi^{\dagger}_{+l\sigma}(\bm{r})\psi^{\dagger}_{-l\sigma}(\bm{r}) \psi_{-l\sigma}(\bm{r})\psi_{+l\sigma}(\bm{r}) \\
&+ V \sum_{l\sigma} \int d\bm{r} \psi^{\dagger}_{+l\sigma}(\bm{r})\psi^{\dagger}_{-l\bar{\sigma}}(\bm{r}) \psi_{-l\bar{\sigma}}(\bm{r})\psi_{+l\sigma}(\bm{r}), 
\end{aligned}
\end{equation}
which includes both intra- ($U$) and inter-sublattice ($V$) interacting strengths.
Both $U$ and $V$ can have important screened Coulomb \cite{Goodwin:2019aa,Pizarro:2019aa,Cea:2021aa}, 
electron-phonon mediated \cite{Wu:2018aa,Choi:2018aa,Lian:2019aa,Choi:2021aa}, and flavor-fluctuation-mediated interaction contributions \cite{Scalapino:2012aa,Fischer:2021aa}.
In the absence of a microscopic theory, we approximate $U$ and $V$ as momentum-independent tunable parameters. 
Our illustrative calculations use $U = -320$ meV$\cdot$nm$^2$ and $V = 480$ meV$\cdot$nm$^2$.  This choice yields 
$T_c\sim2$ K, comparable with experimental observations \cite{Park:2021aa,Hao:2021aa}, in the absence of 
magnetic fields.  No qualitative aspect of our results depends on this model choice.  The dependence of 
$T_c$ in MATBG and MATTG on $\bm{B}_{||}$ is similar for any model with valley-singlet spin-triplet pairing.

\begin{figure}
 \centering
\includegraphics[width=\columnwidth]{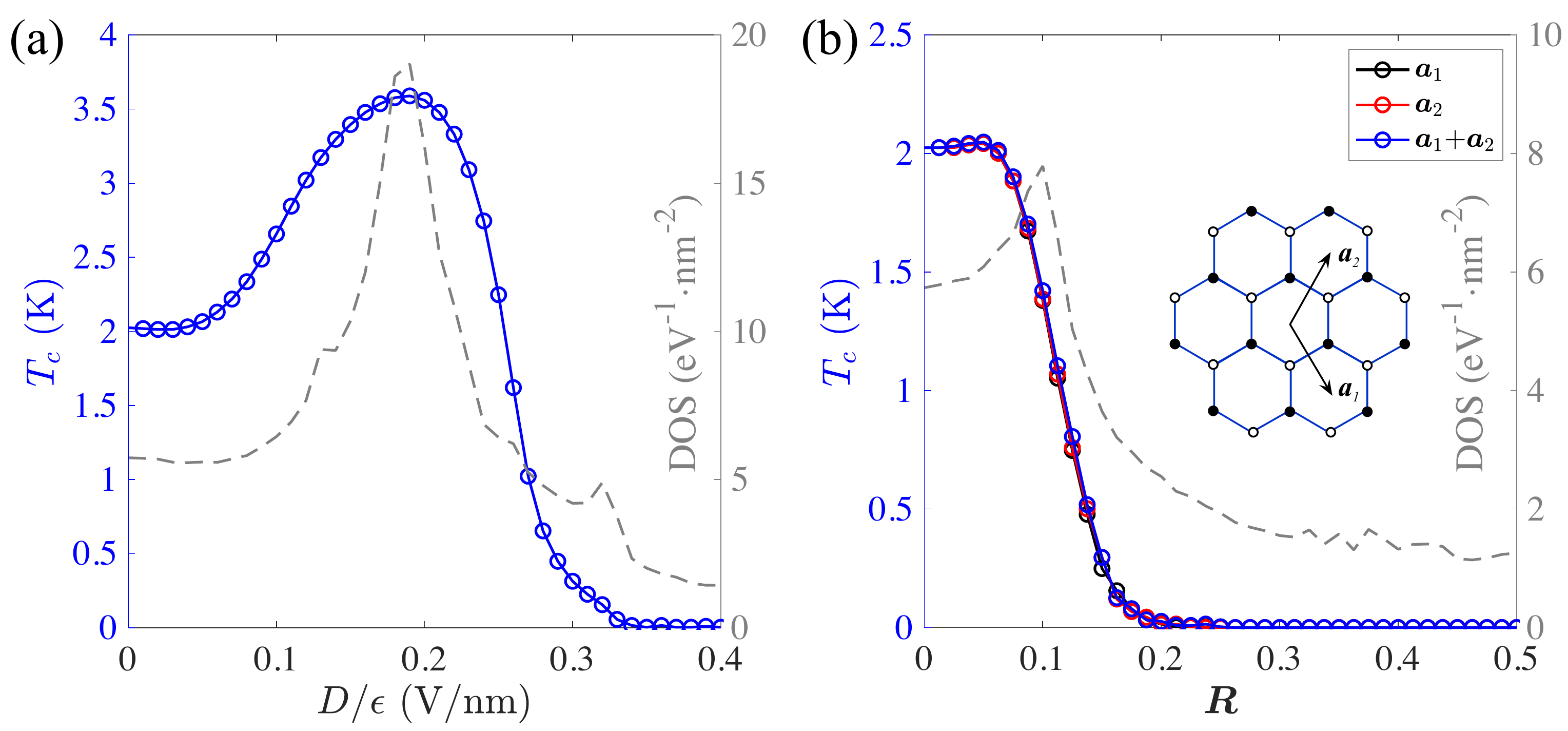}
 \caption{(a) MATTG mean-field critical temperature $T_c$ \textit{vs.} gate electric field $D/\epsilon$. The dashed curve shows the Fermi-level density of states. (b) $T_c$ \textit{vs.} lateral shift $\bm{R}$ of the top or bottom graphene layer. The insert defines the directions of the lateral shift relative to the lattice vectors of the shifted graphene layer. 
 The dashed curve again plots the Fermi level density of states.} 
 \label{fig:figure2}
\end{figure}

\textit{Critical temperature ---} Figure \ref{fig:figure2}(a) shows the MATTG mean-field critical temperature $T_c$ as a 
function of the screened gate electric field $D/\epsilon$.  The gate field breaks $\mathcal{M}_h$ symmetry and 
therefore hybridizes the dispersive Dirac bands and the magic-angle flat bands.
We find that a maximum in $T_c$ is reached at an intermediate value of  
$D/\epsilon \sim 0.19$ V/nm, after which $T_c$ drops, vanishing at $D/\epsilon \sim 0.34$ V/nm. 
The dome-like behavior of $T_c$ {\it vs.} gate field
is due to the change of density of states (DOS) with $D/\epsilon$, which is non-monotonic but 
has a tendency to decrease as $D/\epsilon$ and the mixing between even and odd parity bands it produces increase.
In Fig.~\ref{fig:figure2}(a), the peak of the DOS coincides with the maximum value of $T_c$. 
The dome-like behavior agrees qualitatively with recent experimental observations of $T_c$ in 
MATTG \cite{Park:2021aa,Hao:2021aa}, indicating that DOS variations
likely play a role.
Figure \ref{fig:figure2}(b) illustrates the effect of a lateral shift $\bm{R}$ of the top or bottom 
graphene layer on $T_c$.  Superconductivity is completely suppressed when $\bm{R} \sim 0.2 \bm{a}$ with $\bm{a}$ the lattice vector of the shifted graphene layer. The suppression of $T_c$ is almost isotropic and caused by a dramatic reduction of DOS at sufficiently large lateral shift where the flat bands become more dispersive due to $\mathcal{M}_{h}$ symmetry breaking.  The fact that the critical temperatures are similar in MATTB and MATBG may 
indicate that $\bm{R}$ is small in experimental devices due to energy minimization. 

\begin{figure}
 \centering
\includegraphics[width=1\columnwidth]{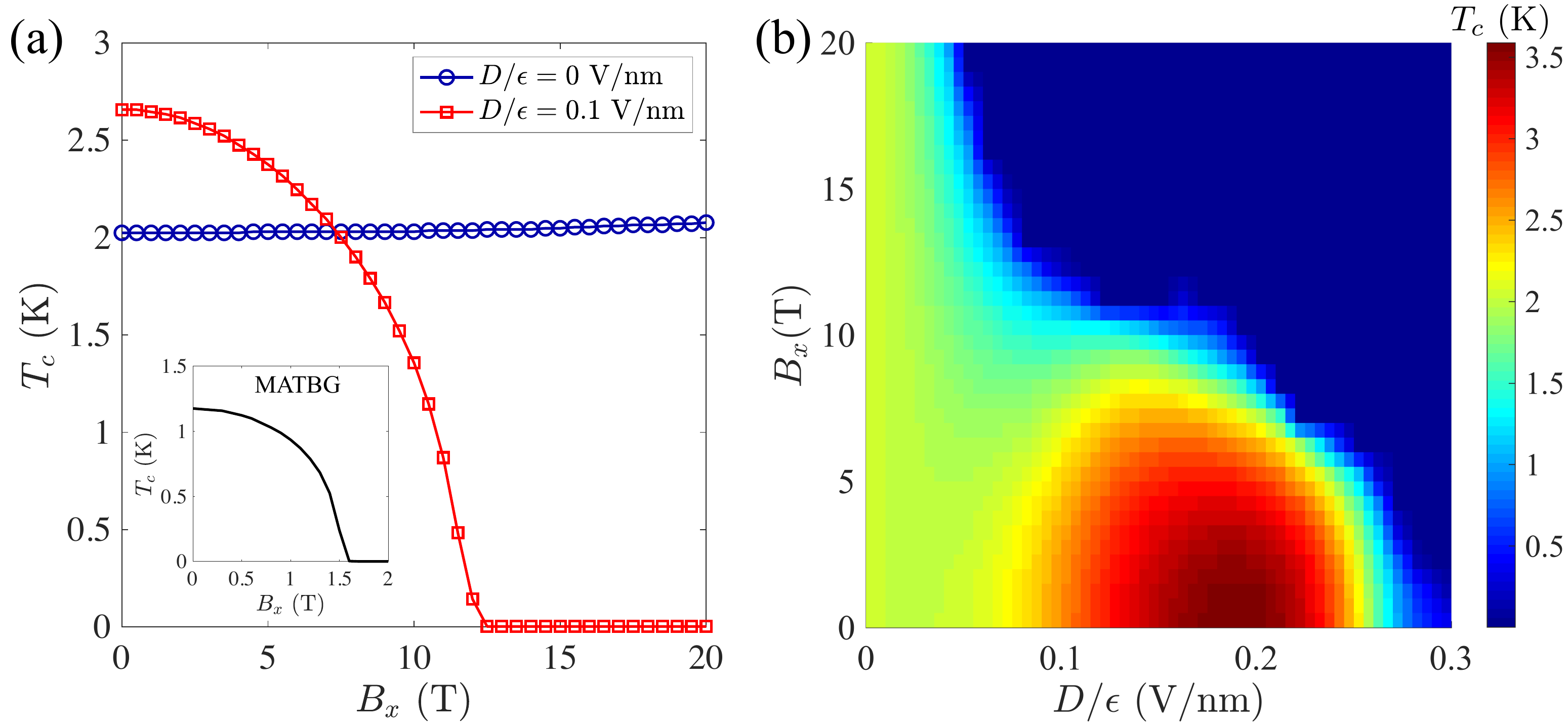}
 \caption{ (a) Critical temperature $T_c$ \textit{vs.} in-plane magnetic field applied along $x$ direction $B_x$ for MATTG at zero and finite gate electric field $D/\epsilon$. The insert plots $T_c$ as function of $B_x$ for MATBG. 
 (b) Two-dimensional color plot of $T_c$ \textit{vs.} $B_x$ and $D/\epsilon$ for MATTG . 
 These results are calculated without lateral shift of the top or bottom graphene layer.}
 \label{fig:figure3}
\end{figure}

\textit{In-plane critical magnetic field ---} Figure~\ref{fig:figure3}(a) shows valley-singlet critical temperatures
$T_c$ calculated at finite values of in-plane magnetic field. Breaking $\mathcal{T}$ symmetry by applying
$B_x$ dramatically suppresses $T_c$ in MATBG superconductor, as illustrated in the insert of Fig.~\ref{fig:figure3}(a). 
The in-plane critical magnetic field obtained for MATBG superconductors is $B_c \sim 1.6$ T, 
which is consistent with experimental observations \cite{Cao:2021ab}. 
Because $B_c$ of MATBG is slightly smaller than the Pauli limit $B_P = (1.86T_c)$ T \cite{Cao:2021ab}, the in-plane magnetic field response 
does not clearly distinguish spin-singlet pairing from spin-triplet pairing. 
In MATTG, on the other hand, $\mathcal{C}_2 \mathcal{M}_h$ or $\mathcal{T} \mathcal{M}_h$ symmetry 
survives in-plane magnetic fields, resulting in the infinite large $B_c$ at zero gate electric field
as illustrated in Fig.~\ref{fig:figure3}(a).  The in-plane magnetic fields actually produce a 
small increase in $T_c$ due to a change in the Fermi level density-of-states.
Application of an gate electric field in MATTG breaks both of $\mathcal{C}_2\mathcal{M}_h$
and $\mathcal{T} \mathcal{M}_h$ symmetries, as summarized in Table~\ref{tab:table1}, leading to the suppression on $T_c$ at finite values of $B_x$ shown in Fig.~\ref{fig:figure3}(a). 
For $D/\epsilon = 0.1$ V/nm, $B_c \sim 12$ T, well in excess of the the Pauli limit $B_P \approx 4.8$ T.
Because orbital pair breaking by $B_x$ is weak at small values of $D/\epsilon $,
studies of $T_c$ {\it vs.} $B_x$ clearly distinguish spin-singlet and spin-triplet superconductivity  in MATTG \cite{Cao:2021aa}. Figure~\ref{fig:figure3}(b) shows $T_c$ {\it vs.} $B_x$ and $D/\epsilon$, illustrating the robustness of 
superconductivity at finite $B_x$ when $D/\epsilon$ is small.  

\begin{figure}
 \centering
\includegraphics[width= 1\columnwidth]{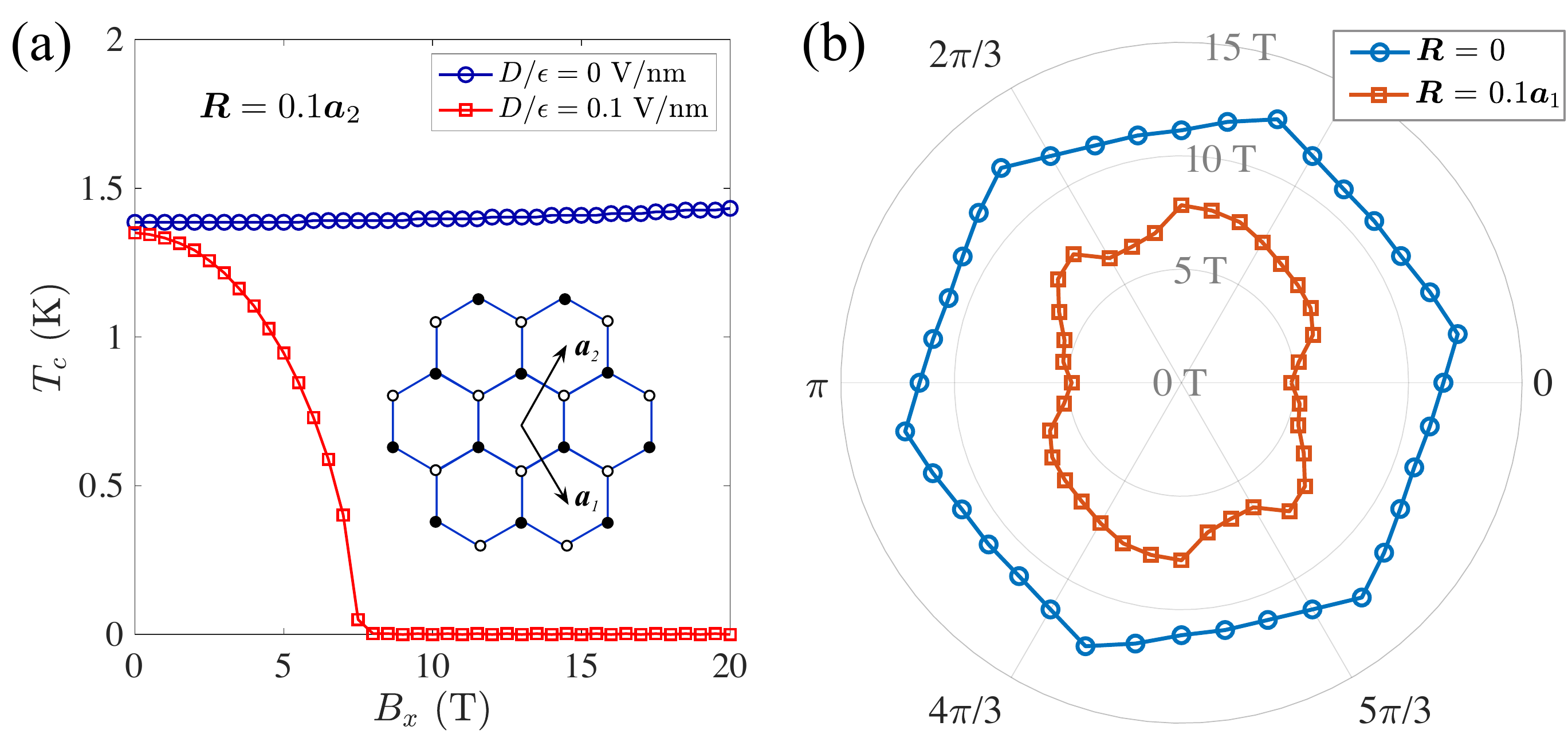}
 \caption{(a) $T_c$ \textit{vs.} $B_x$ and gate electric field at a lateral shift $\bm{R} = 0.1 \bm{a}_2$. (b) Angular dependence of the in-plane critical magnetic field $B_c$ for zero ($\bm{R} =0$) and finite ($\bm{R} = 0.1 \bm{a}_1$) lateral shifts of the top or bottom graphene layer. In (b) we chose $D/\epsilon = 0.1$ V/nm.} 
 \label{fig:figure4}
\end{figure}
 
Figure~\ref{fig:figure4}(a) illustrates the influence of the lateral shift of
top or bottom graphene layer on $T_c$ at finite values of $B_x$.  
As summarized in Table.\ref{tab:table1}, although $\mathcal{M}_h$ and $\mathcal{T}\mathcal{M}_h$ symmetries are broken
by lateral shifts, the $\mathcal{C}_2\mathcal{M}_h$ symmetry survives.  The critical magnetic field 
is therefore infinite at $D/\epsilon =0$ V/nm and large at small values of $D/\epsilon $.
In our model calculations $B_c\sim8$ T for $D/\epsilon = 0.1$ V/nm. 
Figure~\ref{fig:figure4}(b) shows the angle-dependence of $B_c$ for $D/\epsilon = 0.1$ V/nm. 
Note that in the absence of lateral shift, $B_c$ has a sixfold rotation symmetry for valley-singlet pairing, 
even though each valley possesses only $\mathcal{C}_3$ symmetry. A lateral shift of the top or bottom graphene layer 
breaks $\mathcal{C}_3$ symmetry and suppresses $T_c$ as shown in Fig.~\ref{fig:figure2}(b), resulting in a reduced $B_c$ 
with twofold rotation symmetry, as illustrated in Fig.~\ref{fig:figure4}(b). 

 \textit{Discussion ---} The present study provides an explanation for the recent observation of 
 extremely large in-plane critical magnetic field in MATTG superconductors \cite{Cao:2021aa} by 
 relating it to a $\mathcal{C}_2\mathcal{M}_h$ symmetry that survives both in-plane magnetic fields and
 relative lateral shifts of one of the outside graphene layers.  $\mathcal{C}_2\mathcal{M}_h$ symmetry 
 is broken by gate electric fields.
 By combining model band structures with phenomenological electron-electron 
 interactions, we obtain values for $T_c$ at finite $\bm{B}_{||}$ that agree qualitatively with experimental observations, 
 and provide an explanation for the partially contrasting properties of MATBG and MATTG 
 superconductors that is consistent provided that both have spin-triplet, valley-singlet pairing.
 However, we find that in the absence of gate electric fields there is practically no
 suppression of superconductivity by in-plane magnetic fields, whereas the experimental critical fields are finite.
 We attribute this difference to $\mathcal{M}_h$ symmetry-breaking disorder that is always present 
 in realistic material systems, due for example to spatially random vertical electric fields.
 Another likely culprit is random differences between the local twist angle between the top and middle 
 graphene layers and the local twist angle between the bottom and middle graphene layers \cite{Hao:2021aa,Uri:2020aa}.
 Both potential and twist-angle disorder generically break $\mathcal{M}_h$ symmetry 
 and will lead to a finite $B_c$ even if spatially averaged potentials and twist angles preserve this symmetry.
 Finally, $\mathcal{C}_2\mathcal{T}$ symmetry breaking due to Fock self-energies, thought to occur in the insulating state at $\nu=-2$ \cite{Po:2018aa,Xie:2020ab}, would if present suppress superconductivity at zero gate field for non-zero lateral 
 shift $\bm{R}$, as indicated in Table~\ref{tab:table1}. It follows from our analysis that measurements of $B_c$ 
anisotropy can be used to identify $\bm{R}\ne 0$ devices.
 
The model band structures we employ in this work yield a Fermi 
surfaces for $\nu = -2.4$ that has 
two hole pockets centered on $\kappa$ and $\kappa'$ points of the MBZ.
The true character of the Fermi surface underlying the superconducting dome is  
however highly uncertain at present in both MATBG and MATTG, mainly because
the potential of mean-field interaction effects
to alter band dispersion and Fermi surface topology is enlarged
by the narrowness of the single-particle bands \cite{Guinea:2018aa,Xie:2020aa}.
At the mean-field level, the Hartree self energy corrections at negative filling
factors shift the bands around $\kappa$ and $\kappa'$ to 
lower energies relative to bands around $\gamma$, 
providing an opening for hole pockets centered on 
$\gamma$. These electronic structure uncertainties do not alter our main conclusions.

\textit{Acknowledgment ---} The authors acknowledge helpful discussions with Ming Xie and Chunli Huang. This work was supported by DOE grant DE-FG02-02ER45958. Numerical calculations were performed using supercomputing resources at the Texas Advanced Computing Center (TACC).

\end{document}